\documentclass[12pt]{article}
\usepackage{amsmath}
\usepackage{amstext}
\usepackage{amsbsy}
\usepackage{amsfonts}
\usepackage{amssymb}
\usepackage{graphicx}
\usepackage{epsfig}
\usepackage[T1]{fontenc}
\usepackage{palatino}
\usepackage{latexsym}
\usepackage{exscale}
\usepackage{amsmath}
\usepackage{amstext}
\usepackage{amsbsy}
\usepackage{amsfonts}
\usepackage{amssymb}
\usepackage{euscript}
\usepackage{eufrak}
\begin{document}
\include{myinclatex}
\begin{titlepage}
\title{Form Factors of the Nucleon in a Selfconsistent Chiral
Model\thanks{This work is supported by COSY KFA J\"ulich ( 7034833 )}} \author{
C. Arias Tobe\~na\thanks{carlosa@mep.ruhr-uni-bochum.de} and M.F.
Gari\thanks{Manfred.Gari@ruhr-uni-bochum.de}\\Ruhr Universit\"at Bochum
\\Institut f\"ur Theoretische Physik II, Agr. Mittelenergiephysik\\D-44780
Bochum, Germany }  \maketitle
\begin{abstract} In the
present paper we extend the  Schwinger model of the nucleon to the inclusion
of loop corrections. Starting from a chiral model of nucleon and delta for the
mesons $\pi,\rho$ and $A_1$ we extend our nonperturbative sefconsistent
regularisation scheme to the calculation of the  nucleon form factors. Chiral
symmetry requires loops according to the interactions ${\cal L}_{\pi\pi\rho},
{\cal L}_{\rho\rho\rho}, {\cal L}_{\rho\pi A_1}, {\cal L}_{\rho A_1 A_1}$.The
coupling constants are strictly introduced by chiral symmetry. The use of  low
energy theorems and SU(3) relations allow an otherwise parameterfree
calculation of the  form factors of the nucleon.

            \end{abstract}

\end{titlepage}
The determination and understanding of nucleon form factors
 belong to the fundamental problems in hadron physics. As form factors
are dynamical quantities they are exceptionally suited for obtaining
information on the underlying strong interaction of the hadronic
subcomponents. As the nucleon resp. pion are the lightest quark systems
their study is of greatest importance.
 While there is already overwhelming
information on the form factors of the proton there is upcoming information on the
neutron form factors. In the present paper we are interested in a
description of form factors at low momentum transfer i.e. $Q^2 < 0-1 GeV^2$.
            \\
QCD is thought to be the correct theory of the strong forces although several
problems related to its application remain without solution: \begin{itemize}                                                                                                              \item inability to solve dynamics of QCD at hadronic scales because of the large coupling constant.                                                                                       \item natural degrees of freedom of QCD can not be directly related with physical observables.
\end{itemize}                                                                                                                                                                             In the last decade Chiral Perturbation Theory\cite{KOLCK},\cite{WEINBERG} has been applied to many problems in nuclear systems for low-energy regime, namely nucleon-nucleon potential, deuteron properties, few-body forces etc.
Chiral Perturbation Theory combines a chiral lagrangian involving pions, nucleons
and deltas with the underlying symmetries of QCD and a power counting scheme
based on a natural expansion parameter $\frac{Q}{M}$ where Q is the
three-momentum of the particle and M is the nucleon mass. In this approach it
is considered the most general lagrangian which involves pions, nucleons and
deltas and transforms under chiral symmetry as the QCD lagrangian. The
parameters that appear in this lagrangian have to be related to the measured
low-energy quantities. These undetermined coefficients will be fixed by
fitting data from low-energy processes. This approach gets in trouble when the
enery is increased because the number of unknown parameters grows
with the order of the expansion. In the chiral perturbation approach, the
coefficients of the lagrangian are all unknown phenomenological parameters
which have to be determined in a fit to the experiments. The number of unknown
parameters  increase rapidly with increasing precision of the momentum
expansion.  \\
The intention of the present work is to develop an approach within there
is no parameters to play with. Our
aim is to go beyond the  perturbation  approach avoiding the
introduction of new parameters each time we increase the order of the
expansion. A non-perturbative method is developed and used to take into
account higher orders of the expansion.

In the present paper we argue that a "parameter free" calculation
of the electromagnetic and strong formfactors is possible. We mean "parameter
free" in the following sense. Starting from an effective Lagrangian which has
the required invariance properties,such as  gauge invariance , chiral
invariance etc., we aim at a calculation of the currents which involves
otherwise no additional free parameters. There should be no additional cut off
parameters which allow us to "correct" our calculated momentum dependence
of the form factors. We refine the description of refs.\cite{Schwinger}.
Schwinger considers a model consisting of nucleon,$\pi,\rho$ and
$A_1$. The mesons rho and A1 appear as gauge particles. The difficulty in such
an approach is that including particles $\pi,\rho, A_1$ and one must
simultaneously deal with the interactions like  ${\cal L}_{\pi\pi\rho}, {\cal
L}_{\rho\rho\rho}, {\cal L}_{\rho\pi A_1}, {\cal L}_{\rho A_1 A_1}$ (see
also Weinberg \cite{Wein}, Wess and Zumino \cite{Wess} and  Gasiorowicz and
Geffen \cite{Gas}) The calculation of form factors therefore must include these
meson-meson interactions. The consideration of these effects leads directly to
the inclusion of loop corrections. Schwinger did not include loop corrections
in his work as in a perturbation approach the loops are leading to undefined
quantities. Using however  the nonperturbative and
selfconsistent approach of ref.\cite{Flender}, we are able to include these
corrections in a parameterfree way.

We have extented the original Lagrangians of Schwinger  to the inclusion of
delta degrees of freedom which are known to be of considerable importance and
have to be included when we want to compare with experiments.

 Note that the symmetries are
violated only due to the finite mass of the mesons. The imposition of current
field identity on the lagrangian relates the $A_1$ mass to the rho mass, i.e.
  $m_{A_1}={\sqrt2}m_\rho$

\newpage
The interaction terms in the effective Lagrangian are given as follows:\\

nucleon-nucleon-meson \\

${\cal
L}_{NN\pi}=-\frac{g_{NN\pi}}{2M}\bar\psi\tau^i\gamma^{5}\gamma^{\mu}\psi\partial_{\mu}\pi^i$\\
${\cal L}_{NN A_1}=-\frac{g_{NN
A_1}}{2M}\bar\psi\tau^i\gamma^{5}\gamma^{\mu}\psi A^i_1$\\ ${\cal
L}_{NN\rho}=-\frac{g_{\rho}}{2}(\bar\psi\gamma^{\mu}\rho_{\mu}\psi+\frac{\kappa_\rho}{2M}\bar\psi\sigma^{\mu\nu}\partial_\nu\rho_\mu\psi)\tau^i$\\
${\cal
L}_{NN\pi\rho}=+\frac{g_{NN\pi}}{2M}g_{\rho}\bar\psi\gamma_5\gamma^{\mu}\psi\epsilon^{ijl}\tau_i\pi^l\rho_{\mu}^j$\\
\\ \\nucleon-delta-meson and delta-delta-meson \\
\\
${\cal
L}_{N\Delta\pi}=-\frac{g_{N\Delta\pi}}{2M}\bar\Psi_{\mu}\vec\tau_{\Delta
N}(g^{\mu\nu}+\frac{1}{4}\gamma^\mu\gamma^\nu)\Psi\partial_{\nu}\vec\pi$\\
${\cal L}_{N\Delta A_1}=-\frac{g_{N\Delta\pi}
m_{A_1}}{2M}\bar\Psi_{\mu}\vec\tau_{\Delta
N}(g^{\mu\nu}+\frac{1}{4}\gamma^\mu\gamma^\nu)\Psi\vec\pi$\\  ${\cal
L}_{N\Delta\rho}=i\frac{g_{N\Delta\rho}}{2M}\bar\Psi_{\mu}\vec\tau_{\Delta
N}\gamma_\nu\gamma_5\Psi\partial_{\nu}\vec\rho^{\mu\nu}\vec\tau_{\Delta N}$\\
${\cal
L}_{\Delta\Delta\pi}=-\frac{g_{\Delta\Delta\pi}}{2M}\bar\Psi_{\alpha}^{\nu}\vec\tau_{\Delta\Delta}g_{\alpha\beta}\Psi^{\beta}\gamma_\mu\gamma_5\partial_{\nu}\vec\pi$\\ ${\cal L}_{\Delta\Delta A_1}=-\frac{g_{\Delta\Delta A_1} }{2M}\bar\Psi_{\alpha}^{\nu}\vec\tau_{\Delta\Delta}g_{\alpha\beta}\Psi^{\beta}\gamma_\mu\gamma_5\vec A_1^{\mu}$\\  ${\cal
L}_{\Delta\Delta\rho}=g_{\Delta\Delta\rho}\bar\Psi_{\alpha}^{\nu}\vec\tau_{\Delta\Delta}g_{\alpha\beta}\Psi^{\beta}\gamma_\mu\vec\rho_{\nu}$\\
\\
\\  meson-meson-interaction\\
\\
${\cal L}_{\pi\rho
A_1}=-\frac{3}{2}m_{A_1}g_\rho\vec\rho_\mu\cdot(\vec\pi\times\vec A_{1\mu})$\\
${\cal
L}_{\rho\rho\rho}=-{g_{\rho}}\epsilon^{ijk}\rho^k\partial^{\mu}\rho^i\rho_{\mu}^j$\\   ${\cal L}_{\pi\pi\rho}=-g_{\rho}\epsilon^{ijk}\pi^k\partial^{\mu}\pi^i\rho_{\mu}^j$\\  ${\cal L}_{\rho A_1 A_1} =g_\rho (\vec A_{1\nu}\times\partial_\nu\vec A_{1\mu}-\vec A_{1\nu}\times\partial_{\mu}\vec A_{1\nu})$\\    \\
%\end{tabular}
%\end{center}
%\end{table}
%Let us summarize the phenomenology arising from our Lagrangian
%Interaction Lagrangians defined through axial and gauge symmetries
%correspond to the work of Schwinger\cite{}.We extend it this work to the
%inclusion of the delta. Note that the symmetrie is only vileted due to the
%finite mass of the mesons. The free Lagrangian are not shown.The interaction
%agrangian are given in the following  equation 1.
%In the actual calculations we take the leading expressions. The actual
%calculations are not done in a Feynman approach but rather using the unitary
%projection approach by Okubo \cite{oku}   /see/  /Huya
The coupling constants are strictly related by symmetries. The
values  are based on the Goldberger-Treiman relation: $f_\pi
g_{NN\pi}F_{NN\pi}(Q^2=0)=m_n g_A$ and the KSFR formula
${m_\rho}^2=2g_{\rho}^2 {f_\pi}^2$  as well as SU(3) model
predictions. The relation of the different coupling costants are
summarized in table 1.

\begin{table}[h]
\begin{center}
\begin{tabular}[t]{|c||c|}\hline
\bf{}&\bf{}\\ $g_{NN\pi}$                       &
$g_{NN\pi}=\frac{m_{N}g_A}{f_\pi}$\\  \hline
$g_{NN\rho}=\frac{1}{2}g_\rho $
                   & $g_\rho=\frac{m_\rho}{\sqrt(2)
f_\pi}$;  $\kappa_\rho=\kappa^{iv}$\\  \hline  $g_{NN A_1}$&
$g_{NN A_1}= m_{A_1}g_{NN\pi} =m_{A_1}\frac{m_{N}g_A}{f_\pi}$\\
\hline $g_{N\Delta\pi}$  & SU(3)
$g_{N\Delta\pi}=\sqrt{\frac{72}{25}}g_{NN\pi}=\sqrt{\frac{72}{25}}\frac{m_{N}g_A}{f_\pi}$\\
\hline $g_{N\Delta A_1}$   & SU(3) $g_{N\Delta
A_1}=m_{A_1}g_{N\Delta\pi}=m_{A_1}\sqrt{\frac{72}{25}}\frac{m_{N}g_A}{f_\pi}$\\
\hline $g_{N\Delta\rho}$   & SU(3)
$g_{N\Delta\rho}=\frac{g_\rho}{2}(1+\kappa_\rho)\sqrt{\frac{72}{25}}=(1+\kappa_\rho)\sqrt{\frac{72}{25}}\sqrt{2}\frac{m_\rho}{4
f_\pi}$\\  \hline $g_{\Delta\Delta\pi}$                       &
SU(3)
$g_{\Delta\Delta\pi}=\frac{1}{5}g_{NN\pi}=\frac{1}{5}\frac{m_{N}g_A}{f_\pi}$\\
\hline $g_{\Delta\Delta\rho}$                      & SU(3)
$g_{\Delta\Delta\rho}=g_{NN\rho}=\sqrt{2}\frac{m_\rho}{2 f_\pi}$\\
\hline $g_{\Delta\Delta A_1}$                      & SU(3)
$g_{\Delta\Delta
A_1}=m_{A_1}g_{\Delta\Delta\pi}=m_{A_1}\frac{1}{5}\frac{m_{N}g_A}{f_\pi}$\\
\hline
\end{tabular} \end{center} \caption{Summary of the coupling
constants used in the present paper. The values are strictly
related by symmetries  and low energy theorems ( vector and axial
vector gauge, SU(3),KSFR ,Goldberger Treiman and Weinbergs sum
rule. The imposition of current field identity on the broken
symmetry of the Lagrangian leads to $m_{A_1}= \sqrt(2)m_\rho$.
$f_\pi=92.6\pm 0.2 MeV$, $g_A=1.261\pm 0.004$.}
\end{table}

%\clearpage

%\newpage
%In what follows we will represent figure.(\ref{OKULOOP}) as a loop or as a
%triangle diagram depending on where the external source couples.

The electromagnetic (e.m.)isovector current plays a special role through the
relation to the $\rho$-nucleon form factor $F_{NN\rho}$.
The form factors in turn are defined by the current matrixelement
\begin{align}
&<p_f|J_\mu(Q^2)^{(iv)}|p_i>=\\\nonumber
&\chi^{\dagger}\bar u(p_f)\frac{\tau^3}{2}\Bigg\{\gamma_\mu F_1^{(iv)}+\frac{i\sigma_{\mu\nu}q^\nu}{2M}\kappa^{iv}F_2^{(iv)}\Bigg\}u(p_i)\chi
\end{align}
Due to vector dominance we have the following relation between electromagnetic
 and rho form factors:
\begin{align}
G_{\gamma,{\it
magnetic}}^{(iv)}(Q^2)&=F_1^{(iv)}(Q^2)+\kappa^{iv}F_2^{(iv)}(Q^2)\quad
F_\gamma^{(iv)}(Q^2)=\Delta_\rho F_\rho(Q^2)\\\nonumber &\to
G^{(iv)}_{\gamma,{\it magnetic} }(Q^2)=\Delta_\rho G_{\rho,{\it
magnetic}}(Q^2)\label{CFI1} \end{align} where $\Delta_\rho$ is the $\rho$
propagator.

The axialvector form factor is given by:

\begin{equation}
G_A(Q^2)=g_A \Delta_{A_1} F_{NN A_1}(Q^2) ,
\end{equation}
where $\Delta_{A_1}$ is the $A_1$ propagator.
\\
\\
In the description
of the nucleon form factors we consider a one loop topology.
The actual calculation are performed by the method of unitary transformation.
This method has several advantages one of which is the treatment of
renormalization contributions. For details see refs. \cite{OKU},\cite{Huyga}.
In addition we use a nonrelativistic formulation of the effective
Lagrangians. Off-shell effects in the form factors are neglected, as they are
known to be small.

Let us discuss first a system which involves only $\gamma$,
N and $\rho$. The arising meson loops to the e.m.form factor
$F_{NN\gamma}(Q^2)$ include different types of contributions:
\\
i) there is  a N  $\rho$-loop with
a coupling of the photon to the nucleon (L1),
ii) there is a loop arising
from the $\gamma-\rho$ interaction (T1).
iii) In addition we have a
wavefunction reorthonormalization contribution which involves a simple N
$\rho$-loop (R1+R2).Compare from Fig.1.
\\

We realize that a
conventional renormalization procedure will fail to remove the divergencies in
our loop integrations. We regularize the loop integral by replacing each bare
vertex by the sum of all reducible higher order vertex corrections.
Higher orders in the coupling constants are included in ladder approximation by
retaining dressed vertices within a one-loop topology.

\begin{figure}[h]
%\rotatebox{00}{\resizebox{12.5cm}{5.0cm}{\epsfbox{/home/carlosa/latex/articulos/ElectricNeutron/fig/G_to_FormI.eps}}}
\rotatebox{00}{\resizebox{12.5cm}{5.0cm}{\epsfbox{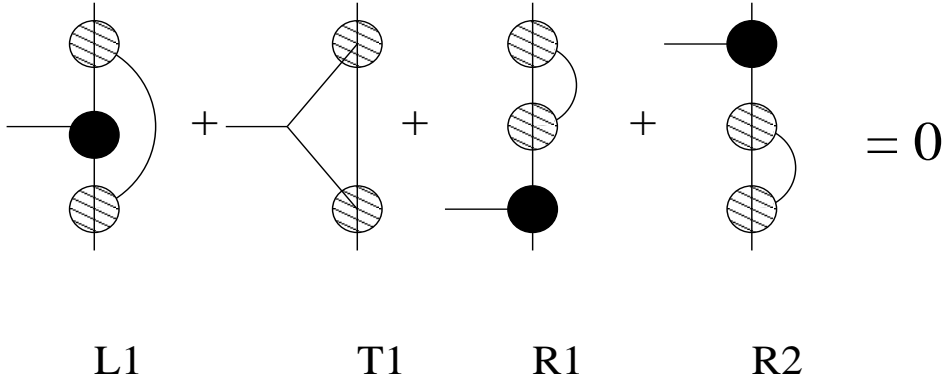}}}

\caption{Illustration of the condition on the e.m. nucleon form factors. This
relation is used to define the form factor. The black circles denote the e.m.
nucleon form factors. The shaded circle denote the $\rho$ nucleon form
factors. Vector dominance leads to  the respective equation for the
$\rho$ nucleon form factor.}\label{GTOFFIGURE}  \end{figure}

For the $\rho$ N system we end up with a condition for the $\gamma$ -nucleon
form factor $F_{NN\gamma}(Q^2)$ as illustrated in Fig.1:

The form factor is determined by the following equation:
\begin{equation}
{F_{NN\gamma}(Q^2)}=-\frac{\Delta_\rho
T1(Q^2,F_{NN\rho}(k^2)F_{NN\rho}({k'}^2)))}
{L1(Q^2,F_{NN\rho}^2(k^2))+R1(Q^2,F_{NN\rho}^2(k^2))+R2(Q^2,F_{NN\rho}^2(k^2))}\label{EQ4}  \end{equation}

This relation tells us that starting the calculations with a form factor which
fullfills the above equation, the meson corrections are included in a one-loop
topology.

Current field identity relates this expression  to the $\rho$ nucleon form
factor $F_{NN\rho}$.   We realize that the $F_{NN\rho}(Q^2)$ obeys a
nonlinear equation of the following form:
\begin{equation}
F_{NN\rho}(Q^2)=-\frac{ T1(Q^2,F_{NN\rho}(k^2)F_{NN\rho}({k'}^2)))}
{L1(Q^2,F_{NN\rho}^2(k^2))+R1(Q^2,F_{NN\rho}^2(k^2))+R2(Q^2,F_{NN\rho}^2(k^2))}\label{EQ5}
\end{equation}

 Actually for a meson baryon
system of N, $\Delta$, $\pi$,$\rho$ and $A_1$ (including
 the interactions $\pi\pi\rho$,$\pi\rho A_1$,$\rho\rho\rho$,
$\rho A_1 A_1$), one has a set of coupled integral equations of the form
equivalent to  equ.(5).  The $F_{NN\rho}$ form factor depends in
the general case on all other  form factors  $F_{NN\pi}$, $F_{NN A_1}$
,$F_{N\Delta\pi}$,$F_{N\Delta\rho}$, ...$F_{\Delta\Delta A_1.}$  Similar
relations hold for the other form factors. In general the form factors F
depend on the form factors on other interactions F and  on form factors
belonging to the same interaction.  Formally this can be written as
\begin{equation}
\begin{split}
&F_{NN\rho}^{(1)}(Q^2)+{\cal
K}^{(1)}_{NN\rho}(Q^2,F_{NN\rho}^{(1)},F_{NN\rho}^{(2)},F_{NN\pi},F_{NN
A_1},..,F_{\Delta\Delta A_1})=0\\ &F_{NN\rho}^{(2)}(Q^2)+{\cal
K}^{(1)}_{NN\rho}(Q^2,F_{NN\rho}^{(1)},F_{NN\rho}^{(2)},F_{NN\pi},F_{NN
A_1},...,F_{\Delta\Delta A_1})=0\\ &F_{NN\pi}(Q^2)+{\cal
K}_{NN\pi}(Q^2,F_{NN\rho}^{(1)},F_{NN\rho}^{(2)},F_{NN\pi},F_{NN
A_1}, ...,F_{\Delta\Delta A_1})=0\\ &F_{NN A_1}(Q^2)+{\cal K}_{NN
A_1}(Q^2,F_{NN\rho}^{(1)},F_{NN\rho}^{(2)},F_{NN\pi},F_{NN
A_1},...,F_{\Delta\Delta A_1})=0\\ &F_{N\Delta\pi}(Q^2)+{\cal
K}_{N\Delta\pi}(Q^2,F_{NN\rho}^{(1)},F_{NN\rho}^{(2)},F_{NN\pi},F_{NN
A_1}, ...,F_{\Delta\Delta A_1})=0\\ &F_{N\Delta A_1}(Q^2)+{\cal
K}_{N\Delta
A_1}(Q^2,F_{NN\rho}^{(1)},F_{NN\rho}^{(2)},F_{NN\pi},F_{NN
A_1},...,F_{\Delta\Delta A_1})=0\\ &   ...
\end{split}
\end{equation}
where $\cal K$ denotes the meson loop contributions. Here we
discuss the expressions for Dirac and Pauli form factors. Actually
the system is solved for $F_1$ and the magnetic form factor $G_m$.

We realize that the solution of the system of coupled equations
provides a physical self-consistant regularization and gives a parameter
free and nonperturbative solutions for the meson nucleon form factors
$F_\alpha$ within a one loop ladder approach.

The coupled system of
nonlinear integral equations for the meson-baryon form factors are solved with
an ansatz of monopol form: $F(Q^2)$=$\frac{\Lambda^2}{\Lambda^2+Q^2}$.
This ansatz turns out to be a very good approximation at low $Q^2$. A monopol
ansatz simplifies the solution of the coupled integral equations and is good
enough for the present purposes. In many cases a parametrization in terms of
the monopol form is desired for comparison. This approximation is also
consistent with the  approximations of Schwinger in deriving the lagrange
functions. The results for the form factors  are summarized in table1.
 Note that for the tensor couplings constants we have the relation:
$\kappa_\rho=\kappa^{(iv)}=3.706$. This is the result of selfconsistency and
the condition on the form factors.
 Not surprising the $\pi$ turns out to give the most
important contribution to all form factors. The contribution of
the $\rho$ is important. The contribution of $A_1$ is generally
small, however, not negligible for the isovector form factor. We
note the similarity between the $\pi$ and the $A_1$ form factor
scales. A closer look at the equations for $\pi$ and $A_1$, tells
us that this is an expected behaviour and reflects the relations
of the  resp. lagrangians.

\begin{table}[h]

\label{tABLE1}
\begin{center}
\begin{tabular}[t]{|c||c|c|}\hline
\bf{Scales}&$\bf{fm^{-1}}$&\bf{GeV}\\ $\Lambda_{NN\pi}$
    & 4.06         & 0.801  \\ \hline $\Lambda_{NN A_1}$
 & 4.10         & 0.809  \\ \hline
$\Lambda_{NN\rho}$\hbox{\tiny Dirac} & 5.95      & 1.170
\\  \hline $\Lambda_{NN\rho}$\hbox{\tiny Magnetic} & 4.410
    & 0.870      \\  \hline
$\Lambda_{N\Delta\rho}$\hbox{\tiny Magnetic} & 4.3      & 0.840
 \\  \hline $\Lambda_{N\Delta\pi}$       & 4.00
      & 0.78   \\  \hline $\Lambda_{N\Delta A_1}$      & 4.10
      & 0.809   \\  \hline
$\Lambda_{\Delta\Delta\rho}$\hbox{\tiny Dirac}   & 4.6      & 0.9
 \\  \hline $\Lambda_{\Delta\Delta\rho}$\hbox{\tiny
Magnetic} & 4.5      & 0.880
\\  \hline $\Lambda_{\Delta\Delta\pi}$       & 4.10         & 0.809
\\  \hline $\Lambda_{\Delta\Delta A_1}$      & 4.20 & 0.82
\\  \hline \end{tabular} \caption{Solutions of the coupled equations
for the system N, $\Delta$, $\pi$, $\rho$, $A_1$. Form factors are
calculated for space-like momentum transfer. They are parametrized
in the form: $F(Q^2)$=$\frac{\Lambda^2}{\Lambda^2+Q^2}$. }

\end{center}

                            \end{table}

It is interesting to compare the present results with some observed
quantities. Among the possible choices there seems one of
special importance. As we have very little information on the e.m. neutron form
factors we discuss that choice as it is in addition the most sensitive
quantity.

\begin{figure}
\begin{center}
%\rotatebox{00}{\resizebox{14.5cm}{7.5cm}{\epsfbox{/home/carlosa/latex/articulos/ElectricNeutron/fig/GE.eps}}}
\rotatebox{00}{\resizebox{14.5cm}{7.5cm}{\epsfbox{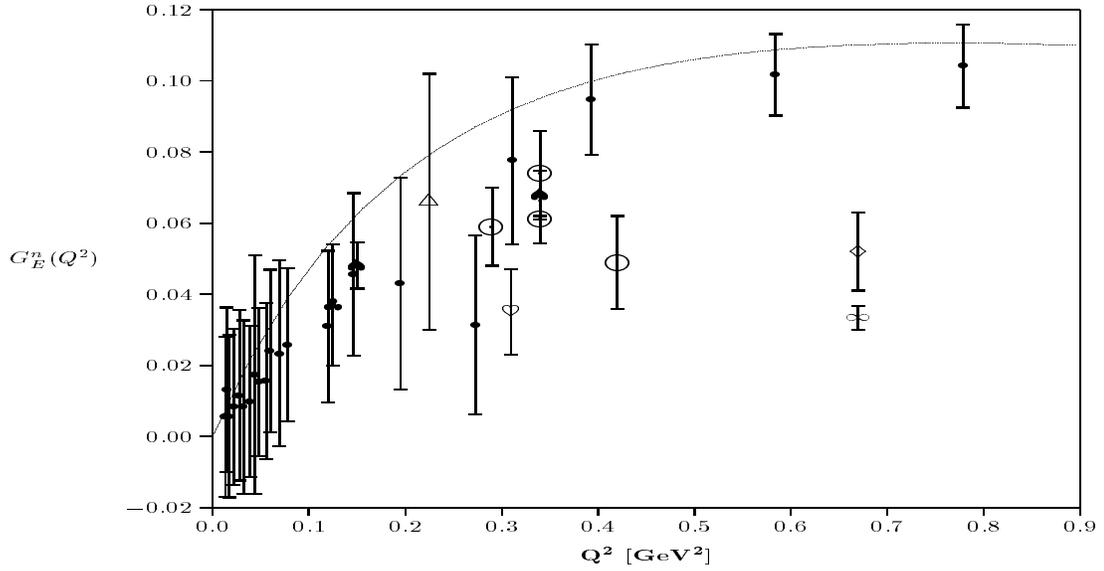}}}
\caption{
Predictions for the electric neutron form factor through the relation to the
measured electric proton form factor. Two choices are shown: a) The dots
("Ruhrdata") are obtained with the use of the measured $G_E^p$  data from the
analysis of \cite{mainz}. b) the line shows the predictions when
the dipole fit i.e.$G_E^p$=$G_{Dipole}$, is used. ( The
derivative of $G_E^n$ at $Q^2=0$ is: $(G_E^n)'$=0.35 ) Experimental
data for the electric neutron form factor are from: $\bigodot$ M.Ostrick {\it
et al}\cite{OSTRICK},$\diamondsuit$ D.Rohe {\it et al}\cite{ROHE}, $\heartsuit$
M.Meyerhoff {\it et al}\cite{MEYERHOFF},$\triangle$ T.Eden {\it et
al}\cite{EDEN},$\clubsuit$ E.Bruins {\it et al}\cite{BRUINS}, $\spadesuit$
C.Herberg {\it et al}\cite{HERBERG}, $\infty$ J.Becker {\it et
al}\cite{BECKER}}\label{GEFFFIGURE}  \end{center}
\end{figure}

\begin{figure}
\begin{center}
%\rotatebox{00}{\resizebox{14.5cm}{6.5cm}{\epsfbox{/home/carlosa/latex/articulos/ElectricNeutron/fig/GM.eps}}}
\rotatebox{00}{\resizebox{14.5cm}{6.5cm}{\epsfbox{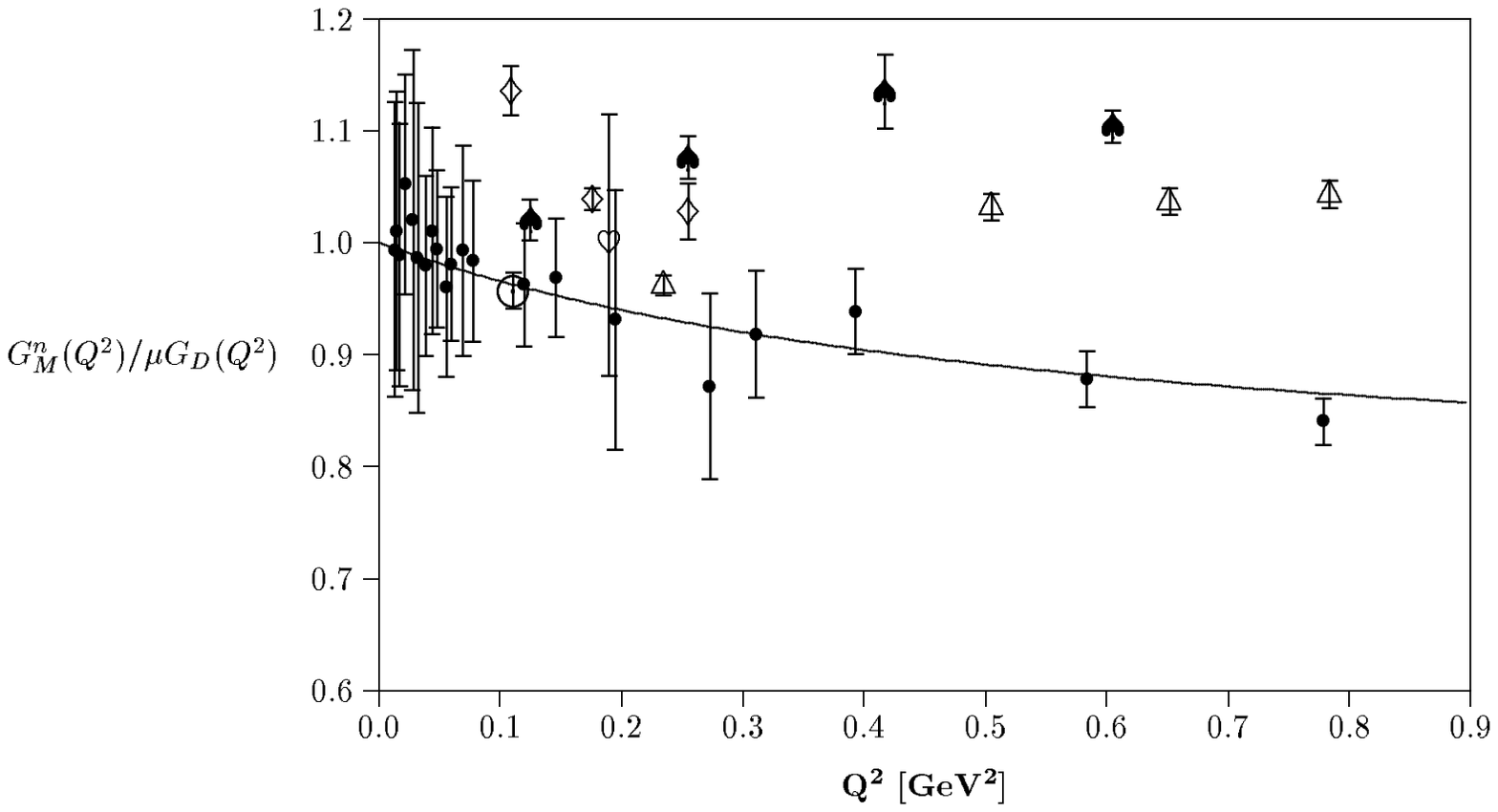}}}

\caption{
Predictions for the magnetic neutron form factor through the relation to the
measured magnetic proton form factor. Two choices are shown: a) The dots
("Ruhrdata") are obtained with the use of the measured $G_M^p$ data
\cite{mainz}. b) the line shows the predictions when we use the dipole
fit i.e. $\frac{G_M^p}{\mu}=G_{Dipole}$, is used. Experimental data for the
magnetic neutron form factor are from:  $\bigodot$H.Anklin {\it et
al}\cite{AKLIN},$\diamondsuit$P.Markowitz {\it et
al}\cite{MARKOWITZ},$\heartsuit$H.Gao {\it et al}\cite{GAO},  $\triangle
$H.Anklin {\it et al}\cite{ANKLINC}, $\spadesuit$E.E.W.Bruins {\it et
al}\cite{BRUINS}}\label{GMFFFIGURE}  \end{center}  \end{figure}

The neutron form factors can be expressed through  measured e.m. form
factors of the proton and the calculated isovector form factors of the nucleon.

\begin{align}
G_E^n(Q^2)&=G_{E({\it
measured})}^p(Q^2)-(F_1^{(iv)}(Q^2)-\frac{Q^2}{4M^2}\kappa^{iv}F_2^{(iv)}(Q^2))\label{C1}\\ G_M^n(Q^2)&=G_{M({\it
measured})}^p(Q^2)-(F_1^{(iv)}(Q^2)+\kappa^{iv}F_2^{(iv)}(Q^2))\label{C2}
\end{align}                       These expressions are used to obtain the
theoretical "Ruhrdata" for the neutron magnetic and electric formfactors.
There are two possibilities in arriving at a description of the neutron form
factors. One uses the widely discussed dipol fit ($G(Q^2)_{Dipole}$=
$\frac{0.71 [GeV^2]}{0.71 [GeV^2]+Q^2})$ of the experimental proton magnetic
and electric formfactors. Another possibility is to use directly the
experimental data. Both possibilities are shown in
Figs.(\ref{GEFFFIGURE},\ref{GMFFFIGURE}) together with the most recent
measurements. Concerning the proton data we use the Mainz analysis
\cite{mainz} of the world data.
We realize that the dipol fit used as representative for the proton data
is different for the electric and magnetic neutron form factor. This is due to
the smallness of the electric neutron form factor at low $Q^2 $.
\\
At momenta  $Q^2$ =0.4 $GeV^2$ one feels already the deviation
from the monopol form from the exact form. This is especially true
for  the Pauli contribution. The electric neutron form factor is
extremely dependent on the shape of the isovector form factor. One
realizes a much stronger dependence of the neutron form factors as
compared to the axial form factor. While the axial form factor is
directly given by the $A_1$ form factor (times $\Delta_{A_1}$ )
the neutron form factors are obtained from a difference of proton
and vector form factors. As already mentioned the axial form
factor is in agreement with the world data in the considered
momentum range.

To summarize we have shown that the extended chiral Schwinger model
($\Delta$, $\pi$, $\rho$, $A_1$, $\gamma$ )  together with the nonperturbative
sefconsistent calculation of the mesonic loop corrections leads to
an interesting description of the nucleon formfactors. As the calculations are
obtained    with no  additional parameters, one expects a similar quality of
the isoscalar current. In this case the model has to be extended to the
inclusion of $\omega$ and $\phi$ as well as strange  particles like K,
$\Lambda$... . We expect no significant changes from the isoscalar
contribution in the coupling to the isovector current. There is  however,
hope that an extension of the present treatment to a calculation of the
nucleon-nucleon potential  will be managable. The use of scales in the
region of the obtained values for the meson nucleon form factors have already
been shown to be possible. \cite{Ruhrpot}.

\end{document}